\documentclass[aps,prm,reprint,superscriptaddress,longbibliography]{revtex4-1}

\usepackage{amsmath,amssymb,graphicx,upgreek}
\usepackage{xcolor}
\usepackage[colorlinks=true,urlcolor=blue,linkcolor=blue,citecolor=blue,bookmarks=false]{hyperref}

\begin{document}
\title{Interplay of itinerant magnetism and reentrant spin-glass behavior in Fe$_{x}$Cr$_{1-x}$}

\author{Georg Benka}
 \email{georg.benka@frm2.tum.de}
 \affiliation{Physik-Department, Technische Universit\"at M\"unchen, D-85748 Garching, Germany}

\author{Andreas Bauer}
 \email{andreas.bauer@ph.tum.de}
 \affiliation{Physik-Department, Technische Universit\"at M\"unchen, D-85748 Garching, Germany}

\author{Philipp Schmakat}
 \affiliation{Physik-Department, Technische Universit\"at M\"unchen, D-85748 Garching, Germany}

\author{Steffen S\"{a}ubert}
 \altaffiliation[Present address: ]{Department of Physics, Colorado State University, Fort Collins, Colorado 80523-1875, USA}
 \affiliation{Physik-Department, Technische Universit\"at M\"unchen, D-85748 Garching, Germany}

\author{Marc Seifert}
 \affiliation{Physik-Department, Technische Universit\"at M\"unchen, D-85748 Garching, Germany}

\author{Pau Jorba}
 \affiliation{Physik-Department, Technische Universit\"at M\"unchen, D-85748 Garching, Germany}

\author{Christian Pfleiderer}
 \affiliation{Physik-Department, Technische Universit\"at M\"unchen, D-85748 Garching, Germany}

\date{\today}

\begin{abstract}
	When suppressing the itinerant antiferromagnetism in chromium by doping with the isostructual itinerant ferromagnet iron, a dome of spin-glass behavior emerges around a putative quantum critical point at an iron concentration $x \approx 0.15$. Here, we report a comprehensive investigation of polycrystalline samples of Fe$_{x}$Cr$_{1-x}$ in the range $0.05 \leq x \leq 0.30$ using x-ray powder diffraction, magnetization, ac susceptibility, and neutron depolarization measurements, complemented by specific heat and electrical resistivity data for $x = 0.15$. Besides antiferromagnetic ($x < 0.15$) and ferromagnetic regimes ($0.15 \leq x$), we identify a dome of reentrant spin-glass behavior at low temperatures for $0.10 \leq x \leq 0.25$ that is preceded by a precursor phenomenon. Neutron depolarization indicates an increase of the size of ferromagnetic clusters with increasing $x$ and the Mydosh parameter $\phi$, inferred from the ac susceptibility, implies a crossover from cluster-glass to superparamagnetic behavior. Taken together, these findings consistently identify Fe$_{x}$Cr$_{1-x}$ as an itinerant-electron system that permits to study the evolution of spin-glass behavior of gradually varying character in unchanged crystalline environment.
\end{abstract}

\maketitle

\section{Motivation}

Chromium is considered as the archetypical itinerant antiferromagnet~\cite{1988_Fawcett_RevModPhys, 1994_Fawcett_RevModPhys}. Interestingly, it shares its body-centered cubic crystal structure $Im\overline{3}m$ with the archetypical itinerant ferromagnet $\alpha$-iron and, at melting temperature, all compositions Fe$_{x}$Cr$_{1-x}$~\cite{2010_Okamoto_Book}. As a result, the Cr--Fe system offers the possibility to study the interplay of two fundamental forms of magnetic order in the same crystallographic environment.

Chromium exhibits transverse spin-density wave order below a N\'{e}el temperature $T_{\mathrm{N}} = 311$~K and longitudinal spin-density wave order below $T_{\mathrm{SF}} = 123$~K~\cite{1988_Fawcett_RevModPhys}. Under substitutional doping with iron, the longitudinal spin-density wave order becomes commensurate at $x = 0.02$. For $0.04 < x$, only commensurate antiferromagnetic order is observed~\cite{1967_Ishikawa_JPhysSocJpn, 1980_Babic_JPhysChemSolids, 1983_Burke_JPhysFMetPhys_I}. The N\'{e}el temperature decreases at first linearly with increasing $x$ and vanishes around $x \approx 0.15$~\cite{1967_Ishikawa_JPhysSocJpn, 1976_Suzuki_JPhysSocJpn, 1978_Burke_JPhysFMetPhys, 1980_Babic_JPhysChemSolids, 1983_Burke_JPhysFMetPhys_I}. Increasing $x$ further, a putative lack of long-range magnetic order~\cite{1978_Burke_JPhysFMetPhys} is followed by the onset of ferromagnetic order at $x \approx 0.18$ with a monotonic increase of the Curie temperature up to $T_{\mathrm{C}} = 1041$~K in pure $\alpha$-iron~\cite{1963_Nevitt_JApplPhys, 1975_Loegel_JPhysFMetPhys, 1980_Fincher_PhysRevLett, 1981_Shapiro_PhysRevB, 1983_Burke_JPhysFMetPhys_II, 1983_Burke_JPhysFMetPhys_III}.

The suppression of magnetic order is reminiscent of quantum critical systems under pressure~\cite{2001_Stewart_RevModPhys, 2007_Lohneysen_RevModPhys, 2008_Broun_NatPhys}, where substitutional doping of chromium with iron decreases the unit cell volume. In comparison to stoichiometric systems tuned by hydrostatic pressure, however, disorder and local strain are expected to play a crucial role in Fe$_{x}$Cr$_{1-x}$. This conjecture is consistent with reports on superparamagnetic behavior for $0.20 \leq x \leq 0.29$~\cite{1975_Loegel_JPhysFMetPhys}, mictomagnetic behavior~\footnote{In mictomagnetic materials, the virgin magnetic curves recorded in magnetization measurements as a function of field lie outside of the hysteresis loops recorded when starting from high field~\cite{1976_Shull_SolidStateCommunications}.} gradually evolving towards ferromagnetism for $0.09 \leq x \leq 0.23$~\cite{1975_Shull_AIPConferenceProceedings}, and spin-glass behavior for $0.14 \leq x \leq 0.19$~\cite{1979_Strom-Olsen_JPhysFMetPhys, 1980_Babic_JPhysChemSolids, 1981_Shapiro_PhysRevB, 1983_Burke_JPhysFMetPhys_I, 1983_Burke_JPhysFMetPhys_II, 1983_Burke_JPhysFMetPhys_III}. 
 
Despite the rather unique combination of properties, notably a metallic spin glass emerging at the border of both itinerant antiferromagnetic and ferromagnetic order, comprehensive studies addressing the magnetic properties of Fe$_{x}$Cr$_{1-x}$ in the concentration range of putative quantum criticality are lacking. In particular, a classification of the spin-glass regime, to the best of our knowledge, has not been addressed before.

Here, we report a study of polycrystalline samples of Fe$_{x}$Cr$_{1-x}$ covering the concentration range $0.05 \leq x \leq 0.30$, i.e., from antiferromagnetic doped chromium well into the ferromagnetically ordered state of doped iron. The compositional phase diagram inferred from magnetization and ac susceptibility measurements is in agreement with previous reports~\cite{1983_Burke_JPhysFMetPhys_I, 1983_Burke_JPhysFMetPhys_II, 1983_Burke_JPhysFMetPhys_III}. As the perhaps most notable new observation, we identify a precursor phenomenon preceding the onset of spin-glass behavior in the imaginary part of the ac susceptibility. For the spin-glass state, analysis of ac susceptibility data recorded at different excitation frequencies by means of the Mydosh parameter, power-law fits, and a Vogel--Fulcher ansatz establishes a crossover from cluster-glass to superparamagnetic behavior as a function of increasing $x$. Microscopic evidence for this evolution is provided by neutron depolarization, indicating an increase of the size of ferromagnetic clusters with $x$.

Our paper is organized as follows. In Sec.~\ref{sec:methods}, the preparation of the samples and their metallurgical characterization by means of x-ray powder diffraction is reported. In addition, experimental details are briefly described. Providing a first point of reference, the presentation of the experimental results starts in Sec.~\ref{sec:results} with the compositional phase diagram as inferred in our study, before turning to a detailed description of the ac susceptibility and magnetization data. Next, neutron depolarization data are presented, allowing to extract the size of ferromagnetically ordered clusters from exponential fits. Exemplary data on the specific heat, electrical resistivity, and high-field magnetization for $x = 0.15$ complete this section. In Sec.~\ref{sec:discussion}, information on the nature of the spin-glass behavior in Fe$_{x}$Cr$_{1-x}$ and its evolution under increasing $x$ is inferred from an analysis of ac susceptibility data recorded at different excitation frequencies. Finally, in Sec.~\ref{sec:conclusion} the central findings of this study are summarized.

\section{Experimental methods}
\label{sec:methods}

Polycrystalline samples of Fe$_{x}$Cr$_{1-x}$ for $0.05 \leq x \leq 0.30$ ($x = 0.05$, 0.10, 0.15, 0.16, 0.17, 0.18, 0.18, 0.19, 0.20, 0.21, 0.22, 0.25, 0.30) were prepared from iron (4N) and chromium (5N) pieces by means of radio-frequency induction melting in a bespoke high-purity furnace~\cite{2016_Bauer_RevSciInstrum}. No losses in weight or signatures of evaporation were observed. In turn, the composition is denoted in terms of the weighed-in amounts of starting material. Prior to the synthesis, the furnace was pumped to ultra-high vacuum and subsequently flooded with 1.4~bar of argon (6N) treated by a point-of-use gas purifier yielding a nominal purity of 9N. For each sample, the starting elements were melted in a water-cooled Hukin crucible and the resulting specimen was kept molten for about 10~min to promote homogenization. Finally, the sample was quenched to room temperature. With this approach, the imminent exsolution of the compound into two phases upon cooling was prevented, as suggested by the binary phase diagram of the Fe--Cr system reported in Ref.~\cite{2010_Okamoto_Book}. From the resulting ingots samples were cut with a diamond wire saw.

\begin{figure}
	\includegraphics[width=1.0\linewidth]{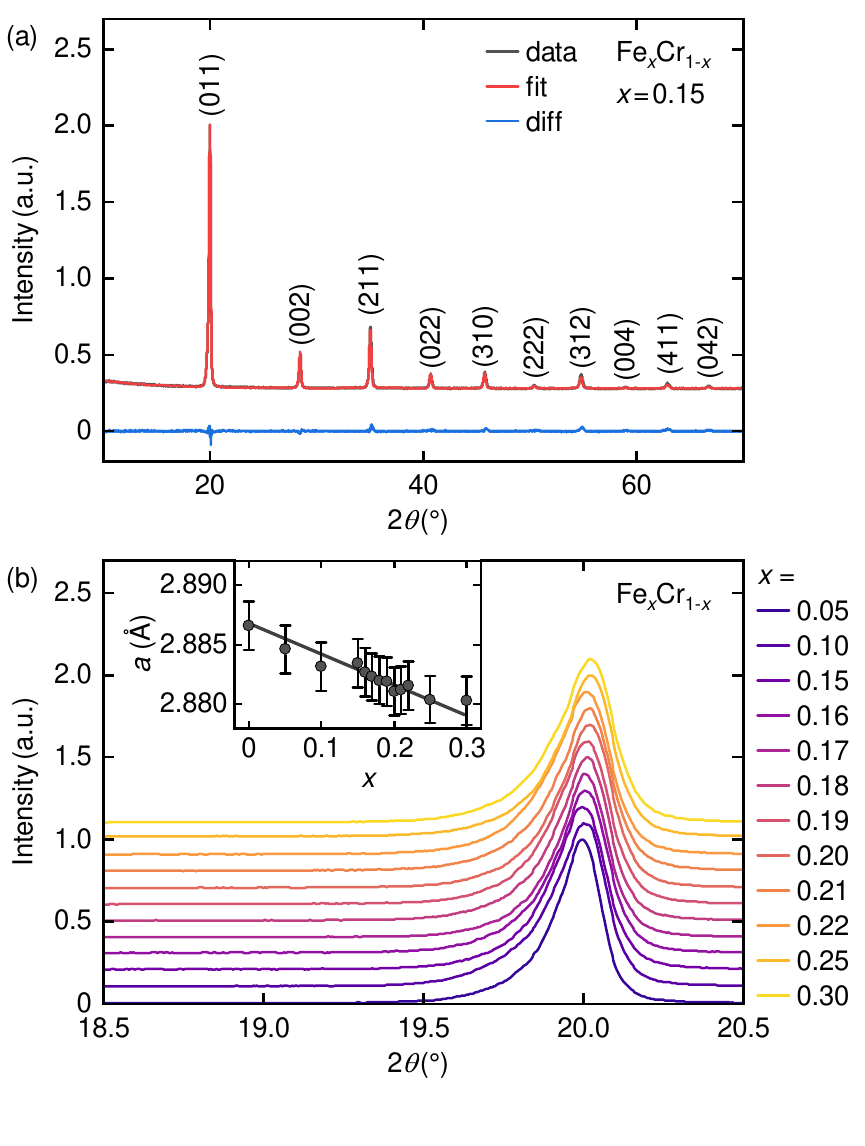}
	\caption{\label{fig:1}X-ray powder diffraction data of Fe$_{x}$Cr$_{1-x}$. (a)~Diffraction pattern for $x = 0.15$. The Rietveld refinement (red curve) is in excellent agreement with the experimental data and confirms the $Im\overline{3}m$ structure. (b)~Diffraction pattern around the (011) peak for all concentrations studied. For clarity, the intensities are normalized and curves are offset by 0.1. Inset: Linear decrease of the lattice constant $a$ with increasing $x$. The solid gray line represents a guide to the eye.}
\end{figure}

Powder was prepared of a small piece of each ingot using an agate mortar. X-ray powder diffraction at room temperature was carried out on a Huber G670 diffractometer using a Guinier geometry. Fig.~\ref{fig:1}(a) shows the diffraction pattern for $x = 0.15$, representing typical data. A Rietveld refinement based on the $Im\overline{3}m$ structure yields a lattice constant $a = 2.883$~\AA.
Refinement and experimental data are in excellent agreement, indicating a high structural quality and homogeneity of the polycrystalline samples. With increasing $x$, the diffraction peaks shift to larger angles, as shown for the (011) peak in Fig.~\ref{fig:1}(b), consistent with a linear decrease of the lattice constant in accordance with Vegard's law.

Measurements of the magnetic properties and neutron depolarization were carried out on thin discs with a thickness of ${\sim}0.5$~mm and a diameter of ${\sim}10$~mm. Specific heat and electrical transport for $x = 0.15$ were measured on a cube of 2~mm edge length and a platelet of dimensions $5\times2\times0.5~\textrm{mm}^{3}$, respectively. 

The magnetic properties, the specific heat, and the electrical resistivity were measured in a Quantum Design physical properties measurement system. The magnetization was measured by means of an extraction technique. If not stated otherwise, the ac susceptibility was measured at an excitation amplitude of 0.1~mT and an excitation frequency of 1~kHz. Additional ac susceptibility data for the analysis of the spin-glass behavior were recorded at frequencies ranging from 10~Hz to 10~kHz. The specific heat was measured using a quasi-adiabatic large-pulse technique with heat pulses of about 30\% of the current temperature~\cite{2013_Bauer_PhysRevLett}. For the measurements of the electrical resistivity the samples were contacted in a four-terminal configuration and a bespoke setup was used based on a lock-in technique at an excitation amplitude of 1~mA and an excitation frequency of 22.08~Hz. Magnetic field and current were applied perpendicular to each other, corresponding to the transverse magneto-resistance.

Neutron depolarization measurements were carried out at the instrument ANTARES~\cite{2015_Schulz_JLarge-ScaleResFacilJLSRF} at the Heinz Maier-Leibniz Zentrum~(MLZ). The incoming neutron beam had a wavelength $\lambda = 4.13$~\AA\ and a wavelength spread $\Delta\lambda / \lambda = 10\%$. It was polarized using V-cavity supermirrors. The beam  was transmitted through the sample and its polarization analyzed using a second polarizing V-cavity. While nonmagnetic samples do not affect the polarization of the neutron beam, the presence of ferromagnetic domains in general results in a precession of the neutron spins. In turn, the transmitted polarization with respect to the polarization axis of the incoming beam is reduced. This effect is referred to as neutron depolarization. Low temperatures and magnetic fields for this experiment were provided by a closed-cycle refrigerator and water-cooled Helmholtz coils, respectively. A small guide field of 0.5~mT was generated by means of permanent magnets. For further information on the neutron depolarization setup, we refer to Refs.~\cite{2015_Schmakat_PhD, 2017_Seifert_JPhysConfSer, 2019_Jorba_JMagnMagnMater}.

All data shown as a function of temperature in this paper were recorded at a fixed magnetic field under increasing temperature. Depending on how the sample was cooled to 2~K prior to the measurement, three temperature versus field histories are distinguished. The sample was either cooled (i)~in zero magnetic field (zero-field cooling, zfc), (ii)~with the field at the value applied during the measurement (field cooling, fc), or (iii)~in a field of 250~mT (high-field cooling, hfc). For the magnetization data as a function of field, the sample was cooled in zero field. Subsequently, data were recorded during the initial increase of the field to $+250$~mT corresponding to a magnetic virgin curve, followed by a decrease to $-250$~mT, and a final increase back to $+250$~mT.

\section{Experimental results}
\label{sec:results}

\subsection{Phase diagram and bulk magnetic properties}

\begin{figure}
	\includegraphics[width=1.0\linewidth]{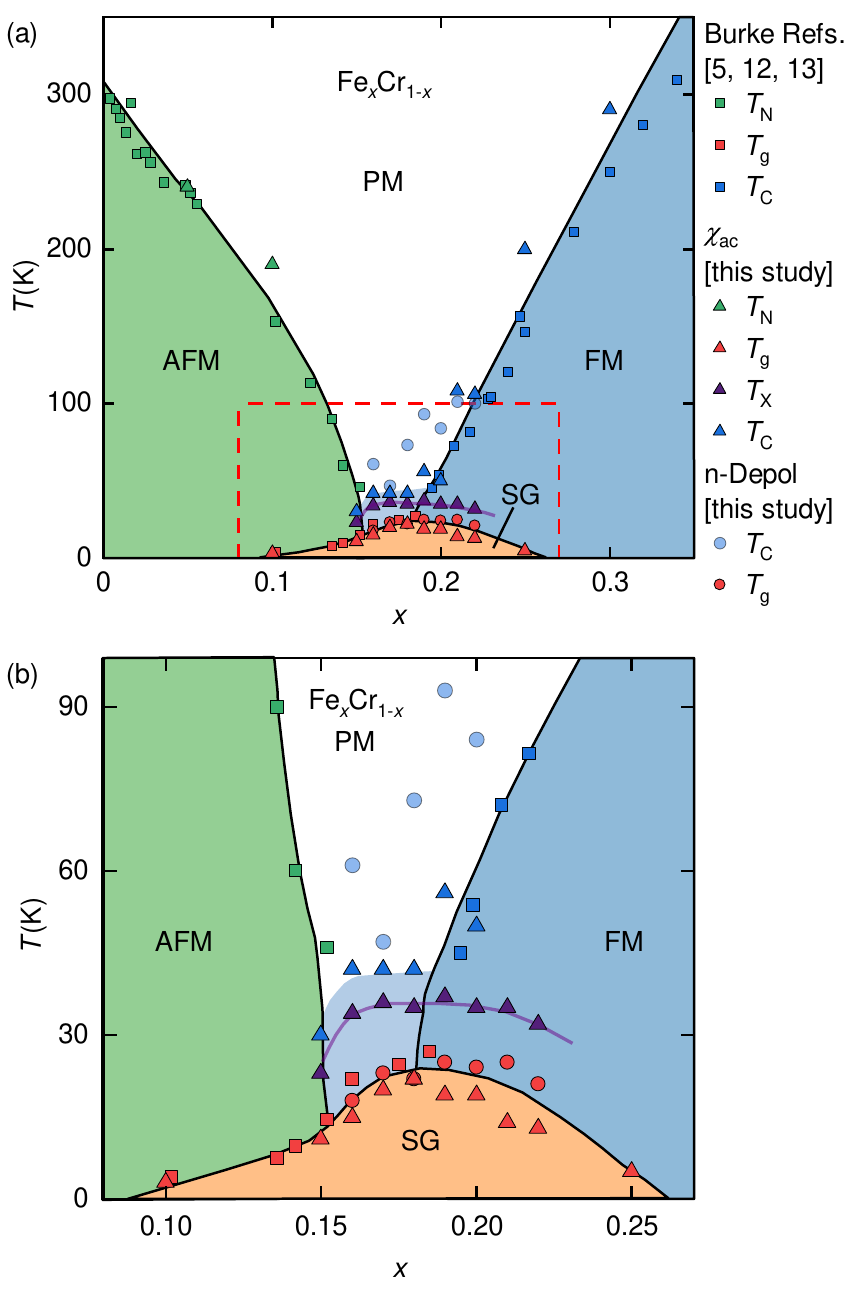}
	\caption{\label{fig:2}Zero-field composition--temperature phase diagram of Fe$_{x}$Cr$_{1-x}$. Data inferred from ac susceptibility, $\chi_{\mathrm{ac}}$, and neutron depolarization are combined with data reported by Burke and coworkers~\cite{1983_Burke_JPhysFMetPhys_I, 1983_Burke_JPhysFMetPhys_II, 1983_Burke_JPhysFMetPhys_III}. Paramagnetic~(PM), antiferromagnetic~(AFM), ferromagnetic~(FM), and spin-glass~(SG) regimes are distinguished. A precursor phenomenon is observed above the dome of spin-glass behavior (purple line). (a)~Overview. (b) Close-up view of the regime of spin-glass behavior as marked by the dashed box in panel (a).}
\end{figure}

The presentation of the experimental results starts with the compositional phase diagram of Fe$_{x}$Cr$_{1-x}$, illustrating central results of our study. An overview of the entire concentration range studied, $0.05 \leq x \leq 0.30$, and a close-up view around the dome of spin-glass behavior are shown in Figs.~\ref{fig:2}(a) and \ref{fig:2}(b), respectively. Characteristic temperatures inferred in this study are complemented by values reported by Burke and coworkers~\cite{1983_Burke_JPhysFMetPhys_I, 1983_Burke_JPhysFMetPhys_II, 1983_Burke_JPhysFMetPhys_III}, in good agreement with our results. Comparing the different physical properties in our study, we find that the imaginary part of the ac susceptibility displays the most pronounced signatures at the various phase transitions and crossovers. Therefore, the imaginary part was used to define the characteristic temperatures as discussed in the following. The same values are then marked in the different physical properties to highlight the consistency with alternative definitions of the characteristic temperatures based on these properties.

Four regimes may be distinguished in the phase diagram, namely paramagnetism at high temperatures (PM, no shading), antiferromagnetic order for small values of $x$ (AFM, green shading), ferromagnetic order for larger values of $x$ (FM, blue shading), and spin-glass behavior at low temperatures (SG, orange shading). We note that faint signatures reminiscent of those attributed to the onset of ferromagnetic order are observed in the susceptibility and neutron depolarization for $0.15 \leq x \leq 0.18$ (light blue shading). In addition, a distinct precursor phenomenon preceding the spin-glass behavior is observed at the temperature $T_{\mathrm{X}}$ (purple line) across a wide concentration range. Before elaborating on the underlying experimental data, we briefly summarize the key characteristics of the different regimes.

We attribute the onset of antiferromagnetic order below the N\'{e}el temperature $T_{\mathrm{N}}$ for $x = 0.05$ and $x = 0.10$ to a sharp kink in the imaginary part of the ac susceptibility, where values of $T_{\mathrm{N}}$ are consistent with previous reports~\cite{1978_Burke_JPhysFMetPhys, 1983_Burke_JPhysFMetPhys_I}. As may be expected, the transition is not sensitive to changes of the magnetic field, excitation frequency, or cooling history. The absolute value of the magnetization is small and it increases essentially linearly as a function of field in the parameter range studied.

We identify the emergence of ferromagnetic order below the Curie temperature $T_{\mathrm{C}}$ for $0.18 \leq x$ from a maximum in the imaginary part of the ac susceptibility that is suppressed in small magnetic fields of a few millitesla. This interpretation is corroborated by the onset of neutron depolarization. The transition is not sensitive to changes of the excitation frequency or cooling history. The magnetic field dependence of the magnetization exhibits a characteristic S-shape with almost vanishing hysteresis, reaching quasi-saturation at small fields. Both characteristics are expected for a soft ferromagnetic material such as iron. For $0.15 \leq x \leq 0.18$, faint signatures reminiscent of those observed for $0.18 \leq x$, such as a small shoulder instead of a maximum in the imaginary part of the ac susceptibility, are interpreted in terms of an incipient onset of ferromagnetic order.

We identify reentrant spin-glass behavior below a freezing temperature $T_{\mathrm{g}}$ for $0.10 \leq x \leq 0.25$ from a pronounced maximum in the imaginary part of the ac susceptibility that is suppressed at intermediate magnetic fields of the order of 50~mT. The transition shifts to lower temperatures with increasing excitation frequency, representing a hallmark of spin glasses. Further key indications for spin-glass behavior below $T_{\mathrm{g}}$ are a branching between different cooling histories in the temperature dependence of the magnetization and neutron depolarization as well as mictomagnetic behavior in the field dependence of the magnetization, i.e., the virgin magnetic curve lies outside the hysteresis loop obtained when starting from high magnetic field.

In addition, we identify a precursor phenomenon preceding the onset of spin-glass behavior at a temperature $T_{\mathrm{X}}$ based on a maximum in the imaginary part of the ac susceptibility that is suppressed in small magnetic fields reminiscent of the ferromagnetic transition. With increasing excitation frequency the maximum shifts to lower temperatures, however at a smaller rate than the freezing temperature $T_{\mathrm{g}}$. Interestingly, the magnetization and neutron depolarization exhibit no signatures at $T_{\mathrm{X}}$.

\subsection{Zero-field ac susceptibility}

\begin{figure}
	\includegraphics[width=1.0\linewidth]{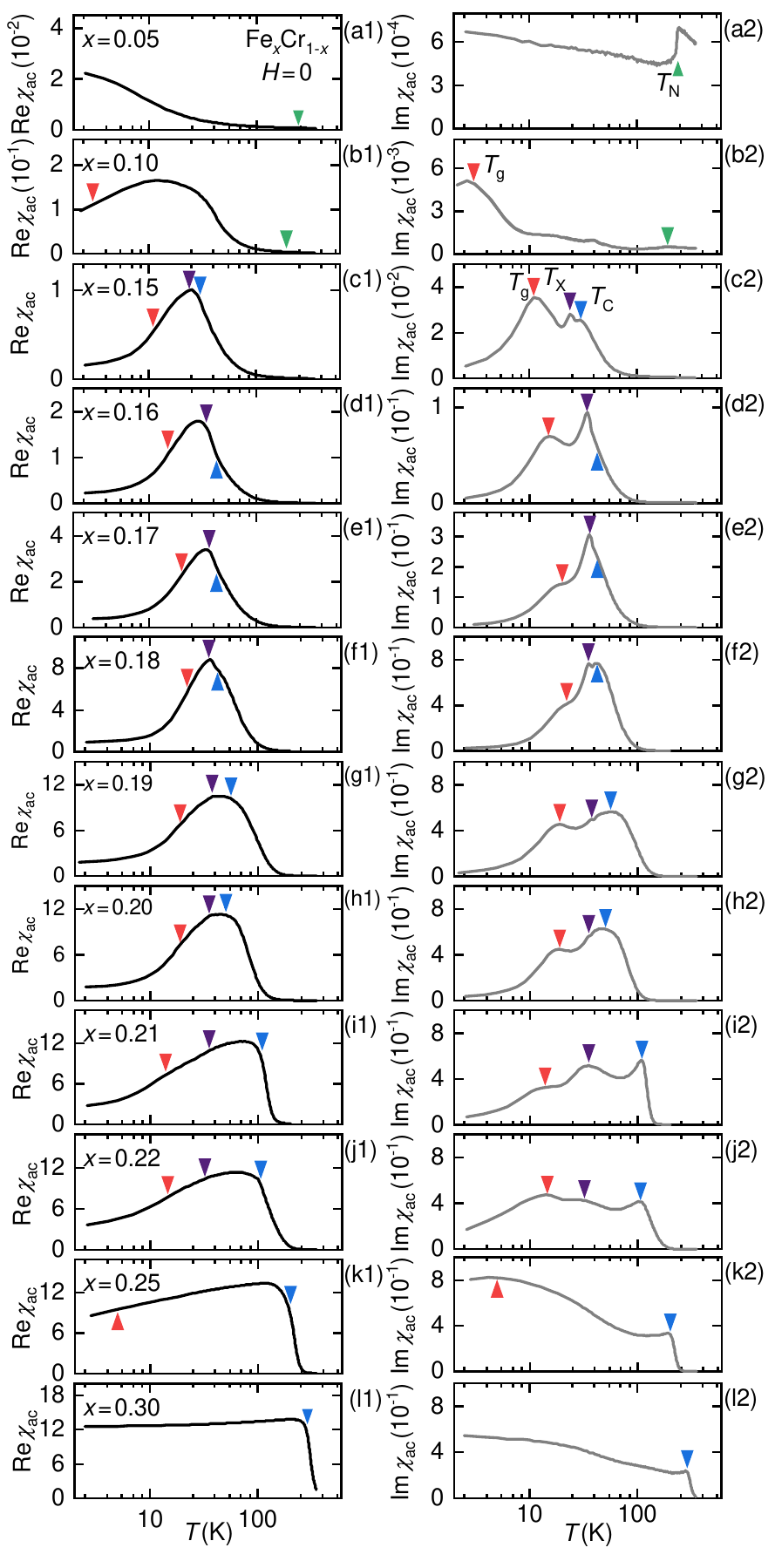}
	\caption{\label{fig:3}Zero-field ac susceptibility as a function of temperature for all samples studied. For each concentration, real part (Re\,$\chi_{\mathrm{ac}}$, left column) and imaginary part (Im\,$\chi_{\mathrm{ac}}$, right column) of the susceptibility are shown. Note the logarithmic temperature scale and the increasing scale on the ordinate with increasing $x$. Triangles mark temperatures associated with the onset of antiferromagnetic order at $T_{\mathrm{N}}$ (green), spin-glass behavior at $T_{\mathrm{g}}$ (red), ferromagnetic order at $T_{\mathrm{C}}$ (blue), and the precursor phenomenon at $T_{\mathrm{X}}$ (purple). The corresponding values are inferred from Im\,$\chi_{\mathrm{ac}}$, see text for details.}
\end{figure}

The real and imaginary parts of the zero-field ac susceptibility on a logarithmic temperature scale are shown in Fig.~\ref{fig:3} for each sample studied. Characteristic temperatures are inferred from the imaginary part and marked by colored triangles in both quantities. While the identification of the underlying transitions and crossovers will be justified further in terms of the dependence of the signatures on magnetic field, excitation frequency, and history, as elaborated below, the corresponding temperatures are referred to as $T_{\mathrm{N}}$, $T_{\mathrm{C}}$, $T_{\mathrm{g}}$, and $T_{\mathrm{X}}$ already in the following.

For small iron concentrations, such as $x = 0.05$ shown in Fig.~\ref{fig:3}(a), the real part is small and essentially featureless, with exception of an increase at low temperatures that may be attributed to the presence of ferromagnetic impurities, i.e., a so-called Curie tail~\cite{1972_DiSalvo_PhysRevB, 2014_Bauer_PhysRevB}. The imaginary part is also small but displays a kink at the N\'{e}el temperature $T_{\mathrm{N}}$. In metallic specimens, such as Fe$_{x}$Cr$_{1-x}$, part of the dissipation detected via the imaginary part of the ac susceptibility arises from the excitation of eddy currents at the surface of the sample. Eddy current losses scale with the resistivity~\cite{1998_Jackson_Book, 1992_Samarappuli_PhysicaCSuperconductivity} and in turn the kink at $T_{\mathrm{N}}$ reflects the distinct change of the electrical resistivity at the onset of long-range antiferromagnetic order.

When increasing the iron concentration to $x = 0.10$, as shown in Fig.~\ref{fig:3}(b), both the real and imaginary parts increase by one order of magnitude. Starting at $x = 0.10$, a broad maximum may be observed in the real part that indicates an onset of magnetic correlations where the lack of further fine structure renders the extraction of more detailed information impossible. In contrast, the imaginary part exhibits several distinct signatures that allow, in combination with data presented below, to infer the phase diagram shown in Fig.~\ref{fig:2}. For $x = 0.10$, in addition to the kink at $T_{\mathrm{N}}$ a maximum may be observed at 3~K which we attribute to the spin freezing at $T_{\mathrm{g}}$.

Further increasing the iron concentration to $x = 0.15$, as shown in Fig.~\ref{fig:3}(c), results again in an increase of both the real and imaginary parts by one order of magnitude. The broad maximum in the real part shifts to slightly larger temperatures. In the imaginary part, two distinct maxima are resolved, accompanied by a shoulder at their high-temperature side. From low to high temperatures, these signatures may be attributed to $T_{\mathrm{g}}$, $T_{\mathrm{X}}$, and a potential onset of ferromagnetism at $T_{\mathrm{C}}$. No signatures related to antiferromagnetism may be discerned. For $x = 0.16$ and 0.17, shown in Figs.~\ref{fig:3}(d) and \ref{fig:3}(e), both the real and imaginary part remain qualitatively unchanged while their absolute values increase further. The characteristic temperatures shift slightly to larger values.

For $x = 0.18$, 0.19, 0.20, 0.21, and 0.22, shown in Figs.~\ref{fig:3}(f)--\ref{fig:3}(j), the size of the real and imaginary parts of the susceptibility remains essentially unchanged. The real part is best described in terms of a broad maximum that becomes increasingly asymmetric as the low-temperature extrapolation of the susceptibility increases with $x$. In the imaginary part, the signature ascribed to the onset of ferromagnetic order at $T_{\mathrm{C}}$ at larger concentrations develops into a clear maximum, overlapping with the maximum at $T_{\mathrm{X}}$ up to $x = 0.20$. For $x = 0.21$ and $x = 0.22$, three well-separated maxima may be attributed to the characteristic temperatures $T_{\mathrm{g}}$, $T_{\mathrm{X}}$, and $T_{\mathrm{C}}$. While both $T_{\mathrm{g}}$ and $T_{\mathrm{X}}$ stay almost constant with increasing $x$, $T_{\mathrm{C}}$ distinctly shifts to higher temperatures.

For $x = 0.25$, shown in Fig.~\ref{fig:3}(k), the signature attributed to $T_{\mathrm{X}}$ has vanished while $T_{\mathrm{g}}$ is suppressed to about 5~K. For $x = 0.30$, shown in Fig.~\ref{fig:3}(l), only the ferromagnetic transition at $T_{\mathrm{C}}$ remains and the susceptibility is essentially constant below $T_{\mathrm{C}}$. Note that the suppression of spin-glass behavior around $x = 0.25$ coincides with the percolation limit of 24.3\% in the crystal structure $Im\overline{3}m$, i.e., the limit above which long-range magnetic order is expected in spin-glass systems~\cite{1978_Mydosh_JournalofMagnetismandMagneticMaterials}. Table~\ref{tab:1} summarizes the characteristic temperatures for all samples studied, including an estimate of the associated errors.

\subsection{Magnetization and ac susceptibility under applied magnetic fields}

\begin{figure*}
	\includegraphics[width=1.0\linewidth]{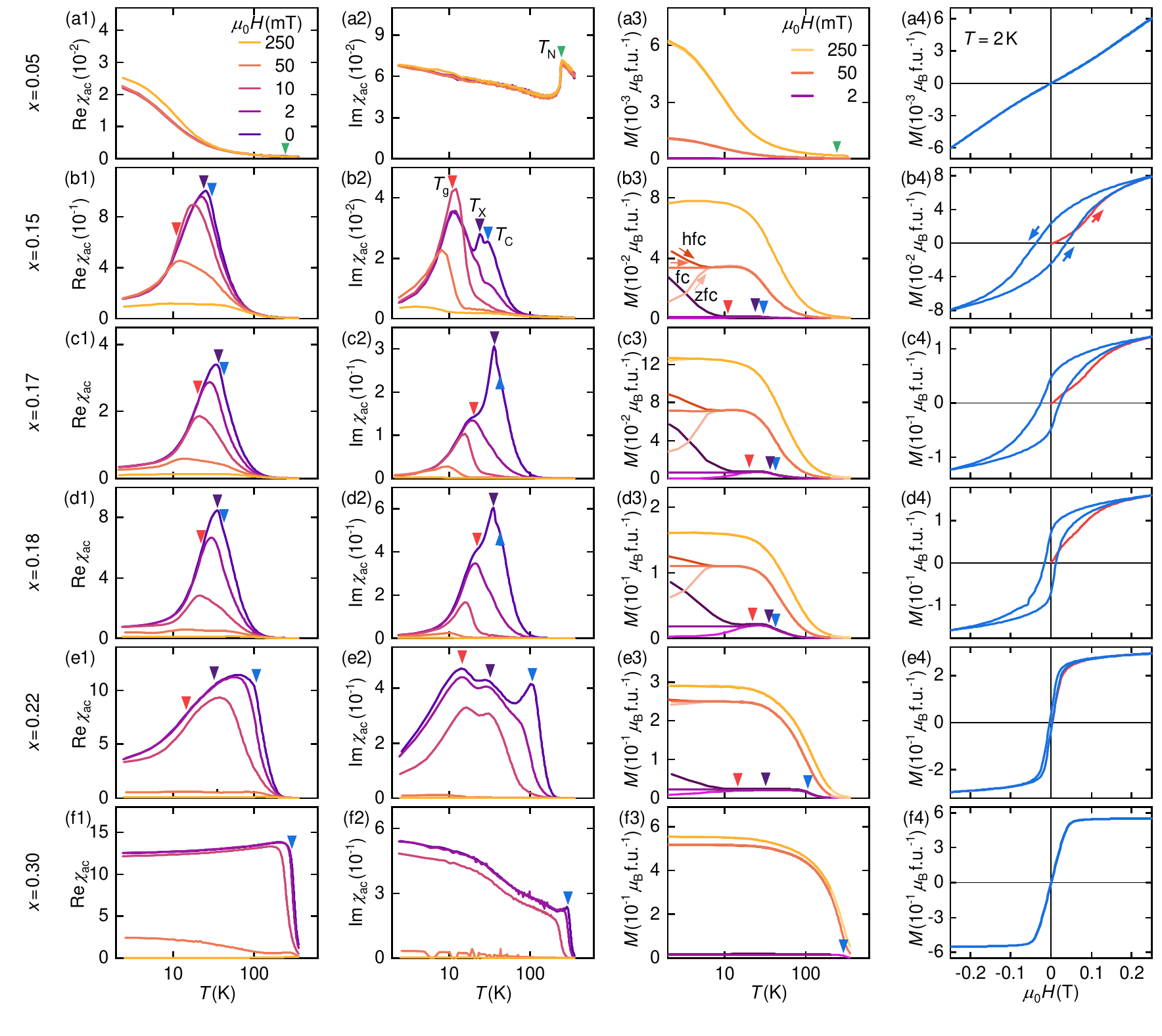}
	\caption{\label{fig:4}Magnetization and ac susceptibility in magnetic fields up to 250~mT for selected concentrations (increasing from top to bottom). Triangles mark the temperatures $T_{\mathrm{N}}$ (green), $T_{\mathrm{g}}$ (red), $T_{\mathrm{C}}$ (blue), and $T_{\mathrm{X}}$ (purple). The values shown in all panels correspond to those inferred from Im\,$\chi_{\mathrm{ac}}$ in zero field. \mbox{(a1)--(f1)}~Real part of the ac susceptibility, Re\,$\chi_{\mathrm{ac}}$, as a function of temperature on a logarithmic scale for different magnetic fields. \mbox{(a2)--(f2)}~Imaginary part of the ac susceptibility, Im\,$\chi_{\mathrm{ac}}$. \mbox{(a3)--(f3)}~Magnetization for three different field histories, namely high-field cooling~(hfc), field cooling (fc), and zero-field cooling (zfc). \mbox{(a4)--(f4)}~Magnetization as a function of field at a temperature of 2~K after initial zero-field cooling. Arrows indicate the sweep directions. The scales of the ordinates for all quantities increase from top to bottom.}
\end{figure*}

In order to justify further the relationship of the signatures in the ac susceptibility with the different phases, their evolution under increasing magnetic field up to 250~mT and their dependence on the cooling history are illustrated in Fig.~\ref{fig:4}. For selected values of $x$, the temperature dependences of the real part of the ac susceptibility, the imaginary part of the ac susceptibility, and the magnetization, shown in the first three columns, are complemented by the magnetic field dependence of the magnetization at low temperature, $T = 2$~K, shown in the fourth column.

For small iron concentrations, such as $x = 0.05$ shown in Figs.~\ref{fig:4}(a1)--\ref{fig:4}(a4), both Re\,$\chi_{\mathrm{ac}}$ and Im\,$\chi_{\mathrm{ac}}$ remain qualitatively unchanged up to the highest fields studied. The associated stability of the transition at $T_{\mathrm{N}}$ under magnetic field represents a key characteristic of itinerant antiferromagnetism, which is also observed in pure chromium. Consistent with this behavior, the magnetization is small and increases essentially linearly in the field range studied. No dependence on the cooling history is observed.

For intermediate iron concentrations, such as $x = 0.15$, $x = 0.17$, and $x = 0.18$ shown in Figs.~\ref{fig:4}(b1) to \ref{fig:4}(d4), the broad maximum in Re\,$\chi_{\mathrm{ac}}$ is suppressed under increasing field. Akin to the situation in zero field, the evolution of the different characteristic temperatures is tracked in Im\,$\chi_{\mathrm{ac}}$. Here, the signatures associated with $T_{\mathrm{X}}$ and $T_{\mathrm{C}}$ proof to be highly sensitive to magnetic fields and are suppressed already above about 2~mT. The maximum associated with the spin freezing at $T_{\mathrm{g}}$ is suppressed at higher field values.

In the magnetization as a function of temperature, shown in Figs.~\ref{fig:4}(b3) to \ref{fig:4}(d3), a branching between different cooling histories may be observed below $T_{\mathrm{g}}$. Compared to data recorded after field cooling (fc), for which the temperature dependence of the magnetization is essentially featureless at $T_{\mathrm{g}}$, the magnetization at low temperatures is reduced for data recorded after zero-field cooling (zfc) and enhanced for data recorded after high-field cooling (hfc). Such a history dependence is typical for spin glasses~\cite{2015_Mydosh_RepProgPhys}, but also observed in materials where the orientation and population of domains with a net magnetic moment plays a role, such as conventional ferromagnets. 

The spin-glass character below $T_{\mathrm{g}}$ is corroborated by the field dependence of the magnetization shown in Figs.~\ref{fig:4}(b4) to \ref{fig:4}(d4), which is perfectly consistent with the temperature dependence. Most notably, in the spin-glass regime at low temperatures, mictomagnetic behavior is observed, i.e., the magnetization of the magnetic virgin state obtained after initial zero-field cooling (red curve) is partly outside the hysteresis loop obtained when starting from the field-polarized state at large fields (blue curves)~\cite{1976_Shull_SolidStateCommunications}. This peculiar behavior is not observed in ferromagnets and represents a hallmark of spin glasses~\cite{1978_Mydosh_JournalofMagnetismandMagneticMaterials}.

For slightly larger iron concentrations, such as $x = 0.22$ shown in Figs.~\ref{fig:4}(e1) to \ref{fig:4}(e4), three maxima at $T_{\mathrm{g}}$, $T_{\mathrm{X}}$, and $T_{\mathrm{C}}$ are clearly separated. With increasing field, first the high-temperature maximum associated with $T_{\mathrm{C}}$ is suppressed, followed by the maxima at $T_{\mathrm{X}}$ and $T_{\mathrm{g}}$. The hysteresis loop at low temperatures is narrower, becoming akin to that of a conventional soft ferromagnet. For large iron concentrations, such as $x = 0.30$ shown in Figs.~\ref{fig:4}(f1) to \ref{fig:4}(f4), the evolution of Re\,$\chi_{\mathrm{ac}}$, Im\,$\chi_{\mathrm{ac}}$, and the magnetization as a function of magnetic field consistently corresponds to that of a conventional soft ferromagnet with a Curie temperature $T_{\mathrm{C}}$ of more than 200~K. For the ferromagnetic state observed here, all domains are aligned in fields exceeding ${\sim}50$~mT. 

\begin{table}	
	\caption{\label{tab:1}Summary of the characteristic temperatures in Fe$_{x}$Cr$_{1-x}$ as inferred from the imaginary part of the ac susceptibility and neutron depolarization data. We distinguish the N\'{e}el temperature $T_{\mathrm{N}}$, the Curie temperature $T_{\mathrm{C}}$, the spin freezing temperature $T_{\mathrm{g}}$, and the precursor phenomenon at $T_{\mathrm{X}}$. Temperatures inferred from neutron depolarization data are denoted with the superscript `D'. For $T_{\mathrm{C}}^{\mathrm{D}}$, the errors were extracted from the fitting procedure (see below), while all other errors correspond to estimates of read-out errors.} 
	\begin{ruledtabular}
	\begin{tabular}{ccccccc}
	$x$ & $T_{\mathrm{N}}$ (K) & $T_{\mathrm{g}}$ (K) & $T_{\mathrm{X}}$ (K) & $T_{\mathrm{C}}$ (K) & $T_{\mathrm{g}}^{\mathrm{D}}$ (K) & $T_{\mathrm{C}}^{\mathrm{D}}$ (K) \\
	\hline
	0.05 & $240 \pm 5$ & -          & -          & -           &-           & -           \\
	0.10 & $190 \pm 5$ & $3 \pm 5$  & -          & -           & -          & -           \\
	0.15 & -           & $11 \pm 2$ & $23 \pm 3$ & $30 \pm 10$ & -          & -           \\
	0.16 & -           & $15 \pm 2$ & $34 \pm 3$ & $42 \pm 10$ & $18 \pm 5$ & $61 \pm 10$ \\
	0.17 & -           & $20 \pm 2$ & $36 \pm 3$ & $42 \pm 10$ & $23 \pm 5$ & $47 \pm 2$  \\
	0.18 & -           & $22 \pm 2$ & $35 \pm 3$ & $42 \pm 10$ & $22 \pm 5$ & $73 \pm 1$  \\
	0.19 & -           & $19 \pm 2$ & $37 \pm 5$ & $56 \pm 10$ & $25 \pm 5$ & $93 \pm 1$  \\
	0.20 & -           & $19 \pm 2$ & $35 \pm 5$ & $50 \pm 10$ & $24 \pm 5$ & $84 \pm 1$  \\
	0.21 & -           & $14 \pm 2$ & $35 \pm 5$ & $108 \pm 5$ & $25 \pm 5$ & $101 \pm 1$ \\
	0.22 & -           & $13 \pm 2$ & $32 \pm 5$ & $106 \pm 5$ & $21 \pm 5$ & $100 \pm 1$ \\
	0.25 & -           &  $5 \pm 5$ & -          & $200 \pm 5$ & -          & -           \\
	0.30 & -           & -          & -          & $290 \pm 5$ & -          & -           \\ 
	\end{tabular}
	\end{ruledtabular}
\end{table}

\subsection{Neutron depolarization}

\begin{figure}
	\includegraphics{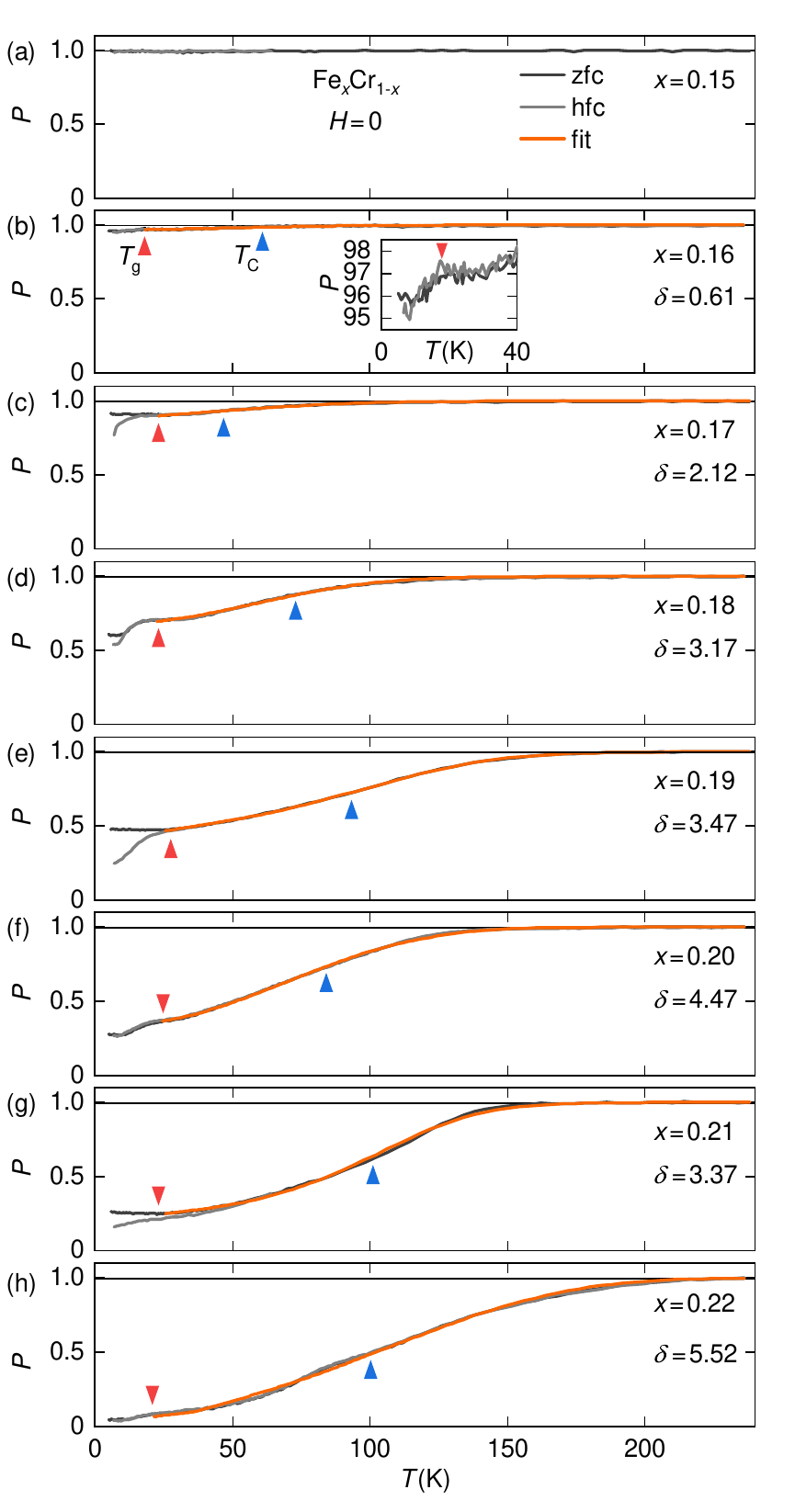}
	\caption{\label{fig:5}Remaining neutron polarization after transmission through 0.5~mm of Fe$_{x}$Cr$_{1-x}$ as a function of temperature for $0.15 \leq x \leq 0.22$ (increasing from top to bottom). Data were measured in zero magnetic field under increasing temperature following initial zero-field cooling (zfc) or high-field cooling (hfc). Colored triangles mark the Curie transition $T_{\mathrm{C}}$ and the freezing temperature $T_{\mathrm{g}}$. Orange solid lines are fits to the experimental data, see text for details.}
\end{figure}

The neutron depolarization of samples in the central composition range $0.15 \leq x \leq 0.22$ was studied to gain further insights on the microscopic nature of the different magnetic states. Figure~\ref{fig:5} shows the polarization, $P$, of the transmitted neutron beam with respect to the polarization axis of the incoming neutron beam as a function of temperature. In the presence of ferromagnetically ordered domains or clusters that are large enough to induce a Larmor precession of the neutron spin during its transit, adjacent neutron trajectories pick up different Larmor phases due to the domain distribution in the sample. When averaged over the pixel size of the detector, this process results in polarization values below 1, also referred to as neutron depolarization. For a pedagogical introduction to the time and space resolution of this technique, we refer to Refs.~\cite{2008_Kardjilov_NatPhys, 2010_Schulz_PhD, 2015_Schmakat_PhD, _Seifert_tobepublished}.

For $x = 0.15$, shown in Fig.~\ref{fig:5}(a), no depolarization is observed. For $x = 0.16$, shown in Fig.~\ref{fig:5}(b), a weak decrease of polarization emerges below a point of inflection at $T_{\mathrm{C}} \approx 60$~K (blue triangle). The value of $T_{\mathrm{C}}$ may be inferred from a fit to the experimental data as described below and is in reasonable agreement with the value inferred from the susceptibility. The partial character of the depolarization, $P \approx 0.96$ in the low-temperature limit, indicates that ferromagnetically ordered domains of sufficient size occupy only a fraction of the sample volume. At lower temperatures, a weak additional change of slope may be attributed to the spin freezing at $T_{\mathrm{g}}$ (red triangle).

For $x = 0.17$, shown in Fig.~\ref{fig:5}(c), both signatures get more pronounced. In particular, data recorded after zero-field cooling (zfc) and high-field cooling (hfc) branch below $T_{\mathrm{g}}$, akin to the branching observed in the magnetization. The underlying dependence of the microscopic magnetic texture on the cooling history is typical for a spin glass. Note that the amount of branching varies from sample to sample. Such pronounced sample dependence is not uncommon in spin-glass systems, though the microscopic origin of these irregularities in Fe$_{x}$Cr$_{1-x}$ remains to be resolved.

When further increasing $x$, shown in Figs.~\ref{fig:5}(c)--\ref{fig:5}(h), the transition temperature $T_{\mathrm{C}}$ shifts to larger values and the depolarization gets more pronounced until essentially reaching $P = 0$ at low temperatures for $x = 0.22$. No qualitative changes are observed around $x = 0.19$, i.e., the composition for which the onset of long-range ferromagnetic order was reported previously~\cite{1983_Burke_JPhysFMetPhys_II}. Instead, the gradual evolution as a function of $x$ suggests that ferromagnetically ordered domains start to emerge already for $x \approx 0.15$ and continuously increase in size and/or number with $x$. This conjecture is also consistent with the appearance of faint signatures in the susceptibility. Note that there are no signatures related to $T_{\mathrm{X}}$.

In order to infer quantitative information, the neutron depolarization data were fitted using the formalism of Halpern and Holstein~\cite{1941_Halpern_PhysRev}. Here, spin-polarized neutrons are considered as they are traveling through a sample with randomly oriented ferromagnet domains. When the rotation of the neutron spin is small for each domain, i.e., when $\omega_{\mathrm{L}}t \ll 2\pi$ with the Larmor frequency $\omega_{\mathrm{L}}$ and the time required for transiting the domain $t$, the temperature dependence of the polarization of the transmitted neutrons may be approximated as  
\begin{equation}\label{equ1}
P(T) = \mathrm{exp}\left[-\frac{1}{3}\gamma^{2}B^{2}_{\mathrm{0}}(T)\frac{d\delta}{v^{2}}\right].
\end{equation}
Here, $\gamma$ is the gyromagnetic ratio of the neutron, $B_{\mathrm{0}}(T)$ is the temperature-dependent average magnetic flux per domain, $d$ is the sample thickness along the flight direction, $\delta$ is the mean magnetic domain size, and $v$ is the speed of the neutrons. In mean-field approximation, the temperature dependence of the magnetic flux per domain is given by 
\begin{equation}\label{equ2}
B_{\mathrm{0}}(T) = {\mu_{0}}^{2} {M_{0}}^{2} \left(1 - \frac{T}{T_{\mathrm{C}}}\right)^{\beta}
\end{equation}
where $\mu_{0}$ is the vacuum permeability, $M_{0}$ is the spontaneous magnetization in each domain, and $\beta$ is the critical exponent. In the following, we use the magnetization value measured at 2~K in a magnetic field of 250~mT as an approximation for $M_{0}$ and set $\beta = 0.5$, i.e., the textbook value for a mean-field ferromagnet. Note that $M_{0}$ more than triples when increasing the iron concentration from $x = 0.15$ to $x = 0.22$, as shown in Tab.~\ref{tab:2}, suggesting that correlations become increasingly important.

Fitting the temperature dependence of the polarization for temperatures above $T_{\mathrm{g}}$ according to Eq.~\eqref{equ1} yields mean values for the Curie temperature $T_{\mathrm{C}}$ and the domain size $\delta$, cf.\ solid orange lines in Fig.~\ref{fig:5} tracking the experimental data. The results of the fitting are summarized in Tab.~\ref{tab:2}. The values of $T_{\mathrm{C}}$ inferred this way are typically slightly higher than those inferred from the ac susceptibility, cf.\ Tab.~\ref{tab:1}. This shift could be related to depolarization caused by slow ferromagnetic fluctuations prevailing at temperatures just above the onset of static magnetic order. Yet, both values of $T_{\mathrm{C}}$ are in reasonable agreement. The mean size of ferromagnetically aligned domains or clusters, $\delta$, increases with increasing $x$, reflecting the increased density of iron atoms. As will be shown below, this general trend is corroborated also by an analysis of the Mydosh parameter indicating that Fe$_{x}$Cr$_{1-x}$ transforms from a cluster glass for small $x$ to a superparamagnet for larger $x$.

\begin{table}
	\caption{\label{tab:2}Summary of the Curie temperature, $T_{\mathrm{C}}$, and the mean domain size, $\delta$, in Fe$_{x}$Cr$_{1-x}$ as inferred from neutron depolarization studies. Also shown is the magnetization measured at a temperature of 2~K in a magnetic field of 250~mT, ${M_{0}}$.}
	\begin{ruledtabular}
	\begin{tabular}{cccc}
	$x$ & $T_{\mathrm{C}}^{\mathrm{D}}$ (K) & $\delta$ ($\upmu$m) & $M_{0}$ ($10^{5}$A/m) \\
	\hline
	0.15 & -           & -               & 0.70 \\
	0.16 & $61 \pm 10$ & $0.61 \pm 0.10$ & 0.84 \\
	0.17 & $47 \pm 2$  & $2.12 \pm 0.15$ & 0.96 \\
	0.18 & $73 \pm 1$  & $3.17 \pm 0.07$ & 1.24 \\
	0.19 & $93 \pm 1$  & $3.47 \pm 0.02$ & 1.64 \\
	0.20 & $84 \pm 1$  & $4.67 \pm 0.03$ & 1.67 \\
	0.21 & $101 \pm 1$ & $3.52 \pm 0.03$ & 2.18 \\
	0.22 & $100 \pm 1$ & $5.76 \pm 0.13$ & 2.27\\
	\end{tabular}
	\end{ruledtabular}
\end{table}

\subsection{Specific heat, high-field magnetometry, and electrical resistivity}

\begin{figure}
	\includegraphics{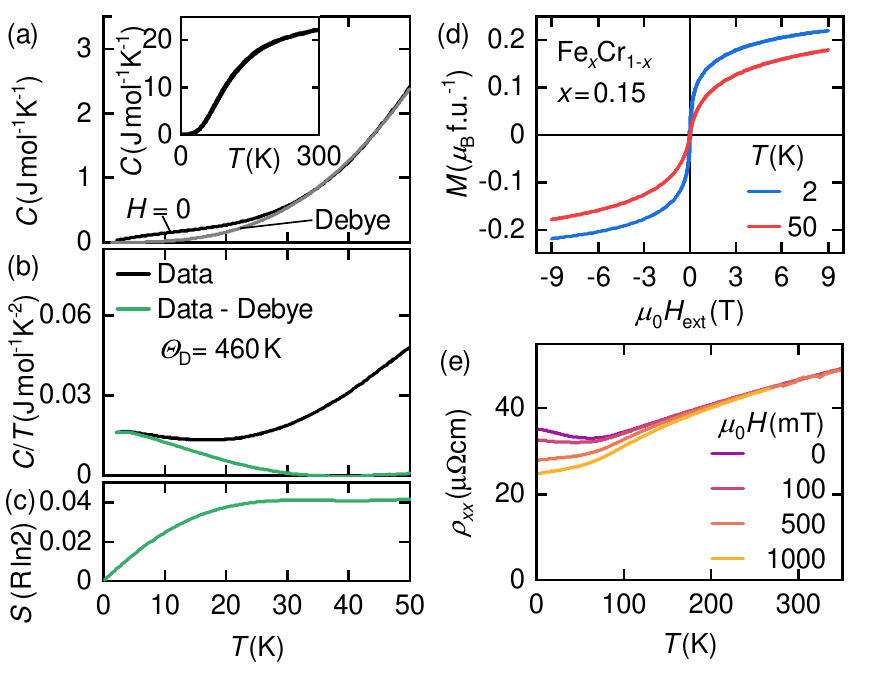}
	\caption{\label{fig:6}Low-temperature properties of Fe$_{x}$Cr$_{1-x}$ with $x = 0.15$. (a)~Specific heat as a function of temperature. Zero-field data (black curve) and an estimate for the phonon contribution using the Debye model (gray curve) are shown. Inset: Specific heat at high temperatures approaching the Dulong--Petit limit. (b)~Specific heat divided by temperature. After subtraction of the phonon contribution, magnetic contributions at low temperatures are observed (green curve). (c)~Magnetic contribution to the entropy obtained by numerical integration. (d)~Magnetization as a function of field up to $\pm9$~T for different temperatures. (e)~Electrical resistivity as a function of temperature for different applied field values.}
\end{figure}

To obtain a complete picture of the low-temperature properties of Fe$_{x}$Cr$_{1-x}$, the magnetic properties at low fields presented so far are complemented by measurements of the specific heat, high-field magnetization, and electrical resistivity on the example of Fe$_{x}$Cr$_{1-x}$ with $x = 0.15$. 

The specific heat as a function of temperature measured in zero magnetic field is shown in Fig.~\ref{fig:6}(a). At high temperatures, the specific heat approaches the Dulong--Petit limit of $C_{\mathrm{DP}} = 3R = 24.9~\mathrm{J}\,\mathrm{mol}^{-1}\mathrm{K}^{-1}$, as illustrated in the inset. With decreasing temperature, the specific heat monotonically decreases, lacking pronounced anomalies at the different characteristic temperatures. 

The specific heat at high temperatures is dominated by the phonon contribution that is described well by a Debye model with a Debye temperature $\mathit{\Theta}_{\mathrm{D}} = 460$~K, which is slightly smaller than the values reported for $\alpha$-iron (477~K) and chromium (606~K)~\cite{2003_Tari_Book}. As shown in terms of the specific heat divided by temperature, $C/T$, in Fig.~\ref{fig:6}(b), the subtraction of this phonon contribution from the measured data highlights the presence of magnetic contributions to the specific heat below ${\sim}$30~K (green curve). As typical for spin-glass systems, no sharp signatures are observed and the total magnetic contribution to the specific heat is rather small~\cite{2015_Mydosh_RepProgPhys}. This finding is substantiated by the entropy $S$ as calculated by means of extrapolating $C/T$ to zero temperature and numerically integrating 
\begin{equation}
S(T) = \int_{0}^{T}\frac{C(T)}{T}\,\mathrm{d}T.
\end{equation}
As shown in Fig.~\ref{fig:6}(c), the magnetic contribution to the entropy released up to 30~K amounts to about $0.04~R\ln2$, which corresponds to a small fraction of the total magnetic moment only.

Insights on the evolution of the magnetic properties under high magnetic fields may be inferred from the magnetization as measured up to $\pm9$~T, shown in Fig.~\ref{fig:6}(d). The magnetization is unsaturated up to the highest fields studied and qualitatively unchanged under increasing temperature, only moderately decreasing in absolute value. The value of 0.22~$\mu_{\mathrm{B}}/\mathrm{f.u.}$ obtained at 2~K and 9~T corresponds to a moment of 1.46~$\mu_{\mathrm{B}}/\mathrm{Fe}$, i.e., the moment per iron atom in Fe$_{x}$Cr$_{1-x}$ with $x = 0.15$ stays below the value of 2.2~$\mu_{\mathrm{B}}/\mathrm{Fe}$ observed in $\alpha$-iron~\cite{2001_Blundell_Book}.

Finally, the electrical resistivity as a function of temperature is shown in Fig.~\ref{fig:6}(e). As typical for a metal, the resistivity is of the order of several ten $\upmu\Omega\,\mathrm{cm}$ and, starting from room temperature, decreases essentially linearly with temperature. However, around 60~K, i.e., well above the onset of magnetic order, a minimum is observed before the resistivity increases towards low temperatures.
Such an incipient divergence of the resistivity with decreasing temperature due to magnetic impurities is reminiscent of single-ion Kondo systems~\cite{1934_deHaas_Physica, 1964_Kondo_ProgTheorPhys, 1987_Lin_PhysRevLett, 2012_Pikul_PhysRevLett}. When magnetic field is applied perpendicular to the current direction, this low-temperature increase is suppressed and a point of inflection emerges around 100~K. This sensitivity with respect to magnetic fields clearly indicates that the additional scattering at low temperatures is of magnetic origin. Qualitatively, the present transport data are in agreement with earlier reports on Fe$_{x}$Cr$_{1-x}$ for $0 \leq x \leq 0.112$~\cite{1966_Arajs_JApplPhys}.

\section{Characterization of the spin-glass behavior}
\label{sec:discussion}

In spin glasses, random site occupancy of magnetic moments, competing interactions, and geometric frustration lead to a collective freezing of the magnetic moments below a freezing temperature $T_{\mathrm{g}}$. The resulting irreversible metastable magnetic state shares many analogies with structural glasses. Depending on the densities of magnetic moments, different types of spin glasses may be distinguished. For small densities, the magnetic properties may be described in terms of single magnetic impurities diluted in a nonmagnetic host, referred to as canonical spin-glass behavior. These systems are characterized by strong interactions and the cooperative spin freezing represents a phase transition. For larger densities, clusters form with local magnetic order and frustration between neighboring clusters, referred to as cluster glass behavior, developing superparamagnetic characteristics as the cluster size increases. In these systems, the inter-cluster interactions are rather weak and the spin freezing takes place in the form of a gradual blocking. When the density of magnetic moments surpasses the percolation limit, long-range magnetic order may be expected.

For compositions close to the percolation limit, so-called reentrant spin-glass behavior may be observed. In such cases, as a function of decreasing temperature first a transition from a paramagnetic to a magnetically ordered state occurs before a spin-glass state emerges at lower temperatures. As both the paramagnetic and the spin-glass state lack long-range magnetic order, the expression ‘reentrant’ alludes to the disappearance of long-range magnetic order after a finite temperature interval and consequently the re-emergence of a state without long-range order~\cite{1993_Mydosh_Book}.

The metastable nature of spin glasses manifests itself in terms of a pronounced history dependence of both microscopic spin arrangement and macroscopic magnetic properties, translating into four key experimental observations; (i) a frequency-dependent shift of the maximum at $T_{\mathrm{g}}$ in the ac susceptibility, (ii) a broad maximum in the specific heat located 20\% to 40\% above $T_{\mathrm{g}}$, (iii) a splitting of the magnetization for different cooling histories, and (iv) a time-dependent creep of the magnetization~\cite{2015_Mydosh_RepProgPhys}. The splitting of the magnetization and the broad signature in the specific heat were addressed in Figs.~\ref{fig:5} and \ref{fig:6}. 

In the following, the frequency dependence of the ac susceptibility will be analyzed by means of three different ways, namely the Mydosh parameter, power law fits, and the Vogel--Fulcher law, permitting to classify the spin-glass behavior in Fe$_{x}$Cr$_{1-x}$ and its change as a function of composition.

\begin{figure}
	\includegraphics[width=0.97\linewidth]{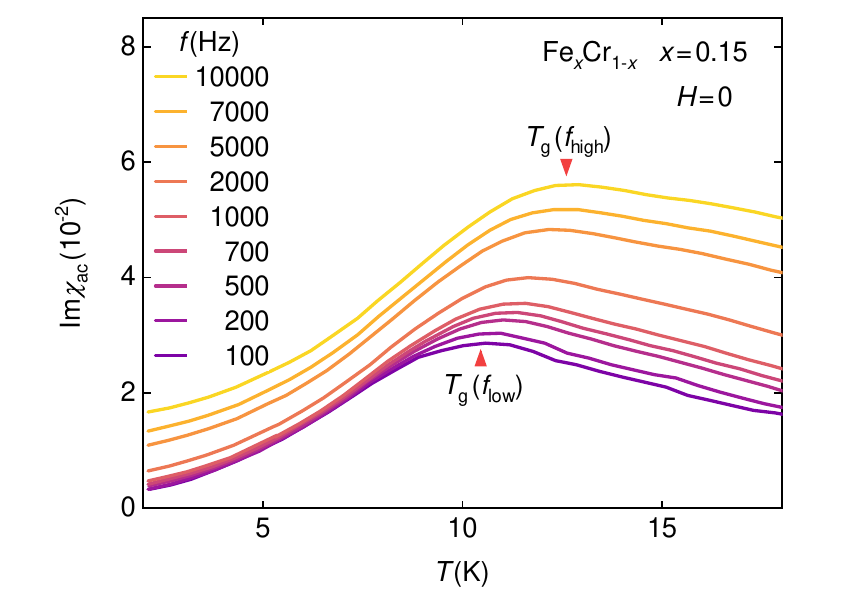}
	\caption{\label{fig:7}Imaginary part of the zero-field ac susceptibility as a function of temperature for Fe$_{x}$Cr$_{1-x}$ with $x = 0.15$ measured at different excitation frequencies $f$. Analysis of the frequency-dependent shift of the spin freezing temperature $T_{\mathrm{g}}$ allows to gain insights on the microscopic nature of the spin-glass state.}
\end{figure}

In the present study, the freezing temperature $T_{\mathrm{g}}$ was inferred from a maximum in the imaginary part of the ac susceptibility as measured at an excitation frequency of 1~kHz. However, in a spin glass the temperature below which spin freezing is observed depends on the excitation frequency $f$, as illustrated in Fig.~\ref{fig:7} for the example of Fe$_{x}$Cr$_{1-x}$ with $x = 0.15$. Under increasing frequency, the imaginary part remains qualitatively unchanged but increases in absolute size and the maximum indicating $T_{\mathrm{g}}$ shifts to higher temperatures. Analyzing this shift in turn provides information on the microscopic nature of the spin-glass behavior. 

The first and perhaps most straightforward approach utilizes the empirical Mydosh parameter $\phi$, defined as 
\begin{equation}
\phi = \left[\frac{T_{\mathrm{g}}(f_{\mathrm{high}})}{T_{\mathrm{g}}(f_{\mathrm{low}})} - 1\right] \left[\ln\left(\frac{f_{\mathrm{high}}}{f_{\mathrm{low}}}\right)\right]^{-1}
\end{equation}
where $T_{\mathrm{g}}(f_{\mathrm{high}})$ and $T_{\mathrm{g}}(f_{\mathrm{low}})$ are the freezing temperatures as experimentally observed at high and low excitation frequencies, $f_{\mathrm{high}}$ and $f_{\mathrm{low}}$, respectively~\cite{1993_Mydosh_Book, 2015_Mydosh_RepProgPhys}. Small shifts associated with Mydosh parameters below 0.01 are typical for canonical spin glasses such as Mn$_{x}$Cu$_{1-x}$, while cluster glasses exhibit intermediate values up to 0.1. Values exceeding 0.1 suggest superparamagnetic behavior~\cite{1993_Mydosh_Book, 2015_Mydosh_RepProgPhys, 1980_Tholence_SolidStateCommun, 1986_Binder_RevModPhys}.

\begin{figure}
	\includegraphics[width=1.0\linewidth]{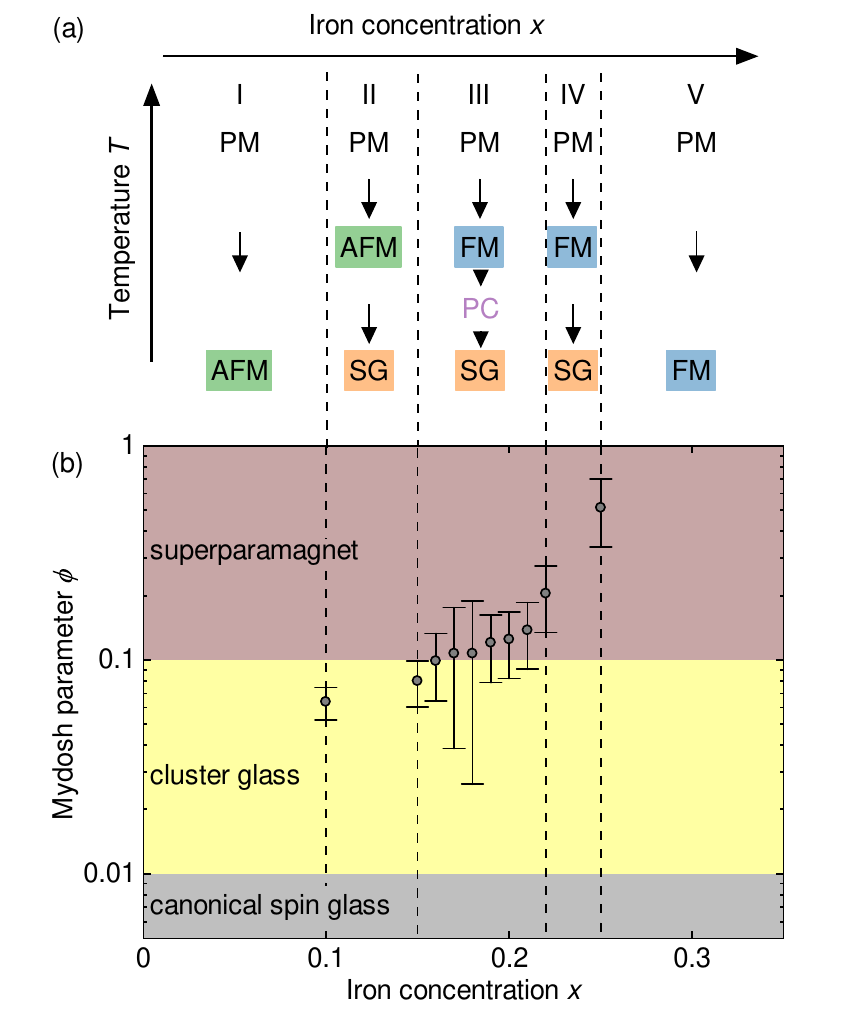}
	\caption{\label{fig:8}Evolution of the Mydosh-parameter in Fe$_{x}$Cr$_{1-x}$. (a)~Schematic depiction of the five different sequences of magnetic regimes observed as a function of temperature for different $x$. The following regimes are distinguished: paramagnetic~(PM), antiferromagnetic~(AFM), ferromagnetic~(FM), spin-glass~(SG). A precursor phenomenon~(PC) may be observed between FM and SG.	(b)~Mydosh parameter $\phi$ as a function of the iron concentration $x$, allowing to classify the spin-glass behavior as canonical ($\phi \leq 0.01$, gray shading), cluster-glass ($0.01 \leq \phi \leq 0.1$, yellow shading), or superparamagnetic ($\phi \geq 0.1$, brown shading). }
\end{figure}

\begin{table*}
	\caption{\label{tab:3}Parameters inferred from the analysis of the spin-glass behavior in Fe$_{x}$Cr$_{1-x}$, namely the Mydosh parameter $\phi$, the zero-frequency extrapolation of the spin freezing temperature $T_\mathrm{g}(0)$, the characteristic relaxation time $\tau_{0}$, the critical exponent $z\nu$, the Vogel--Fulcher temperature $T_{0}$, and the cluster activation energy $E_{a}$. The errors were determined by means of Gaussian error propagation ($\phi$), the distance of neighboring data points ($T_\mathrm{g}(0)$), and statistical deviations of the linear fits ($\tau_{0}$, $z\nu$, $T_{0}$, and $E_{a}$).} 
	\begin{ruledtabular}
		\begin{tabular}{ccccccc}
			$x$ & $\phi$ & $T_\mathrm{g}(0)$ (K) & $\tau_{0}$ ($10^{-6}$~s) & $z\nu$ & $T_{0}$ (K) & $E_{a}$ (K) \\
			\hline
			0.05 & -                 & -              & -               & -             & -              & -              \\
			0.10 & $0.064 \pm 0.011$ & -              & -               & -             & -              & -              \\
			0.15 & $0.080 \pm 0.020$ & $9.1  \pm 0.1$ & $0.16 \pm 0.03$ & $5.0 \pm 0.1$ & $8.5  \pm 0.1$ & $19.9 \pm 0.8$ \\ 
			0.16 & $0.100 \pm 0.034$ & $13.4 \pm 0.1$ & $1.73 \pm 0.15$ & $2.2 \pm 0.0$ & $11.9 \pm 0.1$ & $14.4 \pm 0.3$ \\ 
			0.17 & $0.107 \pm 0.068$ & $18.3 \pm 0.1$ & $6.13 \pm 1.52$ & $1.5 \pm 0.1$ & $16.3 \pm 0.3$ & $12.8 \pm 0.9$ \\ 
			0.18 & $0.108 \pm 0.081$ & $14.5 \pm 0.1$ & $1.18 \pm 0.46$ & $7.0 \pm 0.5$ & $16.9 \pm 0.5$ & $24.2 \pm 2.3$ \\   
			0.19 & $0.120 \pm 0.042$ & $14.2 \pm 0.1$ & $0.47 \pm 0.15$ & $4.5 \pm 0.2$ & $14.6 \pm 0.4$ & $16.3 \pm 1.4$ \\ 
			0.20 & $0.125 \pm 0.043$ & $13.5 \pm 0.1$ & $1.29 \pm 0.34$ & $4.1 \pm 0.2$ & $13.6 \pm 0.3$ & $18.8 \pm 1.3$ \\ 
			0.21 & $0.138 \pm 0.048$ & $9.5  \pm 0.1$ & $1.67 \pm 0.21$ & $4.7 \pm 0.1$ & $10.3 \pm 0.4$ & $12.0 \pm 1.3$ \\ 
			0.22 & $0.204 \pm 0.071$ & $11.7 \pm 0.1$ & $2.95 \pm 0.80$ & $2.6 \pm 0.1$ & $11.3 \pm 0.4$ & $11.3 \pm 1.2$ \\ 
			0.25 & $0.517 \pm 0.180$ & $2.8  \pm 0.1$ & $75.3 \pm 5.34$ & $1.8 \pm 0.1$ & -              & -              \\ 
			0.30 & -                 & -              & -               & -             & -              &                \\ 
		\end{tabular}
	\end{ruledtabular}
\end{table*}

As summarized in Tab.~\ref{tab:3} and illustrated in Fig.~\ref{fig:8}, the Mydosh parameter in Fe$_{x}$Cr$_{1-x}$ monotonically increases as a of function of increasing iron concentration. For small $x$, the values are characteristic of cluster-glass behavior, while for large $x$ they lie well within the regime of superparamagnetic behavior. This evolution reflects the increase of the mean size of ferromagnetic clusters as inferred from the analysis of the neutron depolarization data.

\begin{figure}
	\includegraphics[width=1.0\linewidth]{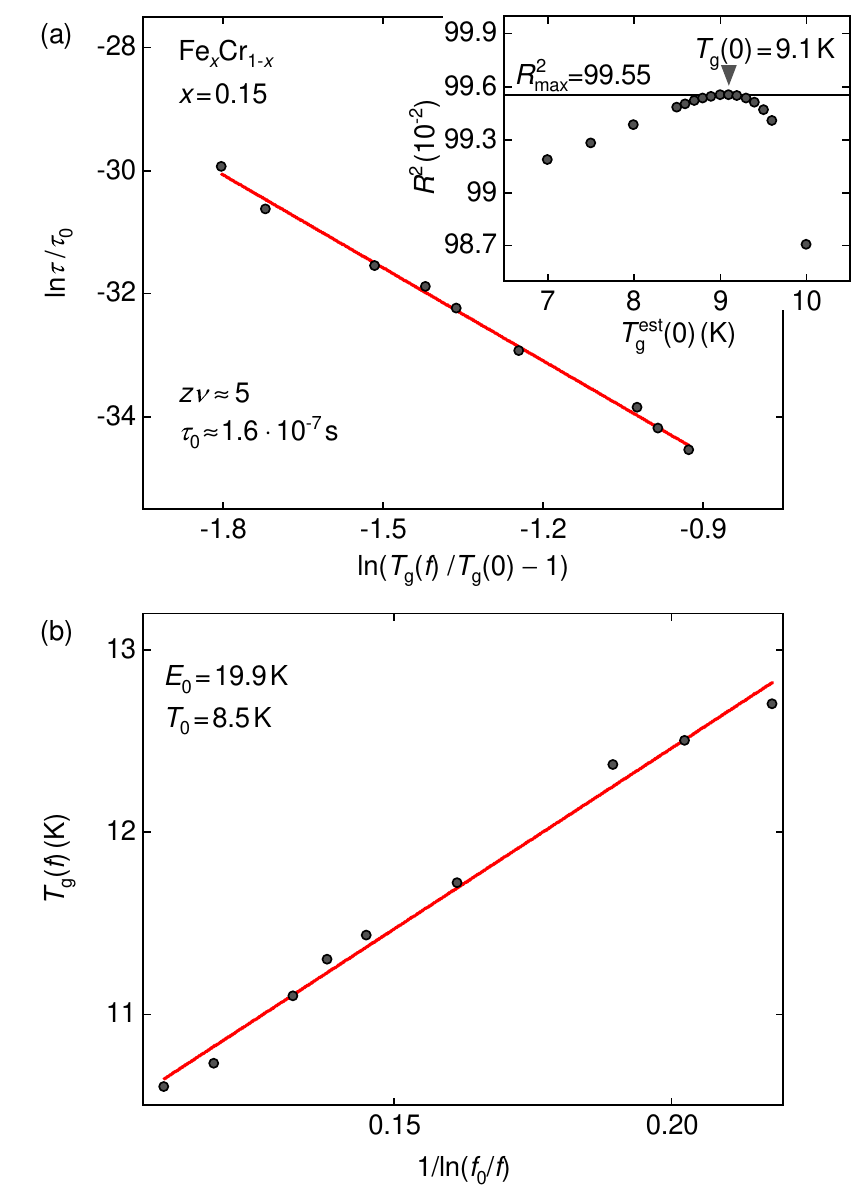}
	\caption{\label{fig:9}Analysis of spin-glass behavior using power law fits and the Vogel--Fulcher law for Fe$_{x}$Cr$_{1-x}$ with $x = 0.15$. (a)~Logarithm of the relaxation time as a function of the logarithm of the normalized shift of the freezing temperature. The red solid line is a power law fit allowing to infer the characteristic relaxation time $\tau_{0}$ and the critical exponent $z\nu$. Inset: Goodness of fit for different estimated zero-frequency extrapolations of the freezing temperature, $T_{\mathrm{g}}^{\mathrm{est}}(0)$. The value $T_{\mathrm{g}}(0)$ used in the main panel is defined as the temperature of highest $R^{2}$. (b)~Spin freezing temperature as a function of the inverse of the logarithm of the ratio of characteristic frequency and excitation frequency. The red solid line is a fit according to the Vogel--Fulcher law allowing to infer the cluster activation energy $E_{a}$ and the Vogel--Fulcher temperature $T_{0}$.}
\end{figure}

The second approach employs the standard theory for dynamical scaling near phase transitions to $T_{\mathrm{g}}$~\cite{1977_Hohenberg_RevModPhys, 1993_Mydosh_Book}. The relaxation time $\tau = \frac{1}{2\pi f}$ is expressed in terms of the power law
\begin{equation}
\tau = \tau_{0} \left[\frac{T_{\mathrm{g}}(f)}{T_{\mathrm{g}}(0)} - 1\right]^{z\nu}
\end{equation}
where $\tau_{0}$ is the characteristic relaxation time of a single moment or cluster, $T_{\mathrm{g}}(0)$ is the zero-frequency limit of the spin freezing temperature, and $z\nu$ is the critical exponent. In the archetypical canonical spin glass Mn$_{x}$Cu$_{1-x}$, one obtains values such as $\tau_{0} = 10^{-13}~\mathrm{s}$, $T_{\mathrm{g}}(0) = 27.5~\mathrm{K}$, and $z\nu = 5$~\cite{1985_Souletie_PhysRevB}.

The corresponding analysis is illustrated in Fig.~\ref{fig:9}(a) for Fe$_{x}$Cr$_{1-x}$ with $x = 0.15$. First the logarithm of the ratio of relaxation time and characteristic relaxation time, $\ln(\frac{\tau}{\tau_{0}})$, is plotted as a function of the logarithm of the normalized shift of the freezing temperature, $\ln\left[\frac{T_{\mathrm{g}}(f)}{T_{\mathrm{g}}(0)} - 1\right]$, for a series of estimated values of the zero-frequency extrapolation $T_{\mathrm{g}}^{\mathrm{est}}(0)$. For each value of $T_{\mathrm{g}}^{\mathrm{est}}(0)$ the data are fitted linearly and the goodness of fit is compared by means of the $R^{2}$ coefficient, cf.\ inset of Fig.~\ref{fig:9}(a). The best approximation for the zero-frequency freezing temperature, $T_{\mathrm{g}}(0)$, is defined as the temperature of highest $R^{2}$. Finally, the characteristic relaxation time $\tau_{0}$ and the critical exponent $z\nu$ are inferred from a linear fit to the experimental data using this value $T_{\mathrm{g}}(0)$, as shown in Fig.~\ref{fig:9}(a) for Fe$_{x}$Cr$_{1-x}$ with $x = 0.15$.

The same analysis was carried out for all compositions Fe$_{x}$Cr$_{1-x}$ featuring spin-glass behavior, yielding the parameters summarized in Tab.~\ref{tab:3}. Characteristic relaxation times of the order of $10^{-6}~\mathrm{s}$ are inferred, i.e., several order of magnitude larger than those observed in canonical spin glasses and consistent with the presence of comparably large magnetic clusters, as may be expected for the large values of $x$. Note that these characteristic times are also distinctly larger than the $10^{-12}~\mathrm{s}$ to $10^{-8}~\mathrm{s}$ that neutrons require to traverse the magnetic clusters in the depolarization experiments. Consequently, the clusters appear quasi-static for the neutron which in turn is a prerequisite for the observation of net depolarization across a macroscopic sample. The critical exponents range from 1.5 to 7.0, i.e., within the range expected for glassy systems~\cite{1980_Tholence_SolidStateCommun, 1985_Souletie_PhysRevB}. The lack of systematic evolution of both $\tau_{0}$ and $z\nu$ as a function of iron concentration $x$ suggests that these parameters in fact may be rather sensitive to details of microscopic structure, potentially varying substantially between individual samples.

The third approach uses the Vogel--Fulcher law, developed to describe the viscosity of supercooled liquids and glasses, to interpret the properties around the spin freezing temperature $T_{\mathrm{g}}$~\cite{1993_Mydosh_Book, 1925_Fulcher_JAmCeramSoc, 1980_Tholence_SolidStateCommun, 2013_Svanidze_PhysRevB}. Calculating the characteristic frequency $f_{0} = \frac{1}{2\pi\tau_{0}}$ from the characteristic relaxation time $\tau_{0}$ as determined above, the Vogel--Fulcher law for the excitation frequency $f$ reads
\begin{equation}
f = f_{0} \exp\left\lbrace-\frac{E_{a}}{k_{\mathrm{B}}[T_{\mathrm{g}}(f)-T_{0}]}\right\rbrace 
\end{equation}
where $k_{\mathrm{B}}$ is the Boltzmann constant, $E_{a}$ is the activation energy for aligning a magnetic cluster by the applied field, and $T_{0}$ is the Vogel--Fulcher temperature providing a measure of the strength of the cluster interactions. As a point of reference, it is interesting to note that values such as $E_{a}/k_{\mathrm{B}} = 11.8~\mathrm{K}$ and $T_{0} = 26.9~\mathrm{K}$ are observed in the archetypical canonical spin glass Mn$_{x}$Cu$_{1-x}$~\cite{1985_Souletie_PhysRevB}.

For each composition Fe$_{x}$Cr$_{1-x}$, the spin freezing temperature $T_{\mathrm{g}}(f)$ is plotted as a function of the inverse of the logarithm of the ratio of characteristic frequency and excitation frequency, $\frac{1}{\ln(f/f_{0})}$, as shown in Fig.~\ref{fig:9}(b) for Fe$_{x}$Cr$_{1-x}$ with $x = 0.15$. A linear fit to the experimental data allows to infer $E_{a}$ and $T_{0}$ from the slope and the intercept. The corresponding values for all compositions Fe$_{x}$Cr$_{1-x}$ featuring spin-glass behavior are summarized in Tab.~\ref{tab:3}. All values of $T_{0}$ and $E_{a}$ are of the order 10~K and positive, indicating the presence of strongly correlated clusters~\cite{2012_Anand_PhysRevB, 2011_Li_ChinesePhysB, 2013_Svanidze_PhysRevB}. Both $T_{0}$ and $E_{a}$ follow roughly the evolution of the spin freezing temperature $T_{\mathrm{g}}$, reaching their maximum values around $x = 0.17$ or $x = 0.18$.

\section{Conclusions}
\label{sec:conclusion}

In summary, a comprehensive study of the magnetic properties of polycrystalline Fe$_{x}$Cr$_{1-x}$ in the composition range $0.05 \leq x \leq 0.30$ was carried out by means of x-ray powder diffraction as well as measurements of the magnetization, ac susceptibility, and neutron depolarization, complemented by specific heat and electrical resistivity data for $x = 0.15$. As our central result, we present a detailed composition--temperature phase diagram based on the combination of a large number of quantities. Under increasing iron concentration $x$, antiferromagnetic order akin to pure Cr is suppressed above $x = 0.15$, followed by the emergence of weak magnetic order developing distinct ferromagnetic character above $x = 0.18$. At low temperatures, a wide dome of reentrant spin-glass behavior is observed for $0.10 \leq x \leq 0.25$, preceded by a precursor phenomenon. Analysis of the neutron depolarization data and the frequency-dependent shift in the ac susceptibility indicate that with increasing $x$ the size of ferromagnetically ordered clusters increases and that the character of the spin-glass behavior changes from a cluster glass to a superparamagnet.

\acknowledgments
We wish to thank P.~B\"{o}ni and S.~Mayr for fruitful discussions and assistance with the experiments. This work has been funded by the Deutsche Forschungsgemeinschaft (DFG, German Research Foundation) under TRR80 (From Electronic Correlations to Functionality, Project No.\ 107745057, Project E1) and the excellence cluster MCQST under Germany's Excellence Strategy EXC-2111 (Project No.\ 390814868). Financial support by the Bundesministerium f\"{u}r Bildung und Forschung (BMBF) through Project No.\ 05K16WO6 as well as by the European Research Council (ERC) through Advanced Grants No.\ 291079 (TOPFIT) and No.\ 788031 (ExQuiSid) is gratefully acknowledged. G.B., P.S., S.S., M.S., and P.J.\ acknowledge financial support through the TUM Graduate School.

\end{document}